\documentclass[12pt]{paper}
\usepackage[pdftex]{graphicx}
\DeclareGraphicsExtensions{.pdf,.jpeg,.png}
\usepackage{amsfonts,amsmath,amssymb}
\usepackage{algorithm,float}
\usepackage[noend]{algpseudocode}
\usepackage{lscape}
\usepackage{flushend}
\usepackage[colorlinks=green,citecolor=blue,urlcolor=blue]{hyperref}
\newcommand{\ABc}{\texttt{ABC}}
\newcommand{\ABcRej}{\texttt{ABC-Rej}}
\newcommand{\ABcRW}{\texttt{ABC-RW}}
\newcommand{\ABcPT}{\texttt{ABC-PT}}

\newcommand{\bfv}{\bf v}
\newcommand{\bfx}{\bf x}

\newcommand{\bfc}{\bf c}

\begin{document}

\title{Approximate Bayesian algorithms for multiple target tracking with binary sensors}

\author{Adrien Ickowicz
\thanks{CSIRO Risk Analytics,Locked Bag 17, North Ryde, NSW, 1670,Australia.}
}

\maketitle


\begin{abstract}
\noindent In this paper, we propose an approximate Bayesian computation approach to perform a multiple target tracking within a binary sensor network. The nature of the binary sensors (\emph{getting closer - moving away} information) do not allow the use of the classical tools (e.g. Kalman Filter, Particle Filer), because the exact likelihood is intractable. To overcome this, we use the particular feature of the likelihood-free algorithms to produce an efficient multiple target tracking methodology. 
\end{abstract}

\begin{keywords}
approximate Bayesian computation, multiple target tracking, binary sensors.
\end{keywords}


%
\section{Introduction}

\subsection{Context}

Sensor networks are systems made of many small and simple sensors deployed over an area in an attempt to sense events of interest within that particular area. In general, the sensors have limited capacities in terms of say range, precision, etc. The ultimate information level for a sensor is a binary one, referring to its output. However, it is important to make a distinction according to the nature of this binary information. Actually, it can be related to a $0-1$ information (non-detection or detection) or to relative $\{ +, -\}$ motion information. For example, if the sensors are getting sound levels, instead of using the real sound level (which may cause confusion between loud near objects and quieter close objects), the sensor may simply report whether the Doppler frequency is suddenly changing, which can be easily translated in whether the target is getting closer or moving away. Moreover, low-power sensors with limited computation and communication capabilities can only perform binary detection. We could also cite video sensors, with the intuitive reasoning: the target is getting closer if its size is increasing. The need to use that kind of sensor networks leads to the development of a model for target tracking in binary sensor networks.

\subsection{Related Works}

The very first work on this kind of binary directional sensors has been made by Aslam et al \cite{Aslam2003}. In their article, the authors provide a modified version of the particle filter to address the issue. They use the geometrical properties implied by such sensors to build a pseudo-likelihood and use it in the filter. However, the approach is limited by the same geometrical constraints the use, and while the velocity can be estimated with a reasonable accuracy, the position estimation is not improved.  In \cite{Ickowicz2008a, Ickowicz2014a}, we developed a new approach for the estimation of both the velocity and the position by learning the target's behaviour. We then use this knowledge to improve the estimation of the position, which leads to a final improvement of both estimation.\\
Several issues appear as soon as someone wants to perform a tracking of multiple targets with a sensor network, even if the sensor measurements are perfect. Indeed, tracking a target involve associating temporal measurements. The problem of associating measurements from sensors to the appropriate tracks, especially when missing data, unknown number of targets, and false observations are present, has been treated \cite{Reid1979}, \cite{YaakovBar-Shalom}, \cite{Blackman2004} using approaches for extracting the most probable hypothesis from a set containing multiple hypotheses all compatible with the actual observations. Some recent work on binary proximity sensors in presented in \cite{Zhu2012}.\\
The extension of the binary problem to multiple target tracking is quite hard due to the fact that by not being able to associate the observations to the targets, the problem moved from a binary issue to a binomial one (instead of having $0-1$ values, we have $0-N_t$ values, where $N_t$ stands for the number of targets). In \cite{Ickowicz2013} we presented a technique based on the sampling of binary matrix and their evaluation through a likelihood function. Then, by applying the Importance Sampling framework, we have been able to achieve an estimation of both position and velocity of multiple targets. 

\subsection{Contribution}

In this article, we change the point of view presented in our previous work \cite{Ickowicz2013}. In that precedent paper, we aimed at generating a draw of sensor observations, $S_t$, and calculate its likelihood, considered as a product of Bernoulli distribution. That approach, while exact (in term the sense that the likelihood calculated was exact and not approximated) was heavily computer demanding, especially if the number of sensors and/or targets is reasonably important. In the present work, however, we consider the tracking in a binary sensor network as an inverse problem, such that by applying a model $\mathcal{M}$ to the velocity and position vectors, we have an output $S_t$, which corresponds to the binary observations by the sensors. Therefore, our aim is to draw the positions and velocities from a well-chosen distribution, apply $\mathcal{M}$, and then evaluate the output by comparison to the observations.\\
The problem with that approach is that the likelihood of the observations/parameters is hard to calculate when the number of target is bigger than $1$. Therefore, we use a likelihood-free approach, first by defining a simple rejection algorithm based on the Euclidean distance, and then by calculating a more complex pseudo-likelihood and computing a random walk Metropolis-Hastings algorithm.\\
The paper is organised as follows. First, we describe the general binary target tracking context, and present the state equations. Then, after having quickly introduced the \ABc~principles, we present the algorithms that we'll use to estimate the tracks of the targets. We finally present some conclusive simulation results, where we compare the performances of the algorithms in function of a number of varying parameters such as the duration of the tracking, the number of targets and the number of sensors in the area.

\section{General Binary Target Tracking Framework}

The targets are assumed to evolve according a general Markov process, described by,
\begin{equation}
\label{eq:traj}
\left\{
\begin{array}{lll}
{\bfv}_t \vert {\bfv}_{t-1} & \sim & \mathcal{N}(F_t {\bfv}_{t-1}, Q_t) \\
{\bfx}_t & = & {\bfx}_{t-1} + {\bfv}_{t-1}
\end{array}
\right.
\end{equation}
for ${t}=1,2...$ where $\mathcal{N}(\mu, \sigma^2)$ is a Gaussian distribution with mean $\mu$ and variance $\sigma^2$. This modelling allows the target to change its trajectory or to maintain a general direction, e.g. we will talk about a low-manoeuvring (respectively highly-manoeuvring) target when $\sigma^2$ is small (resp. big). However, for the sake of simplicity, we will focus on the special case $F_{t} = I$ and ${\bf Q}_t = \sigma^2 I$, that is, the targets are independent. Finally, the starting position is assumed to be unknown, as well as the initial velocity vector.

\begin{figure}[ht!]
\centering
{\includegraphics[width=7.5cm, height=6cm]{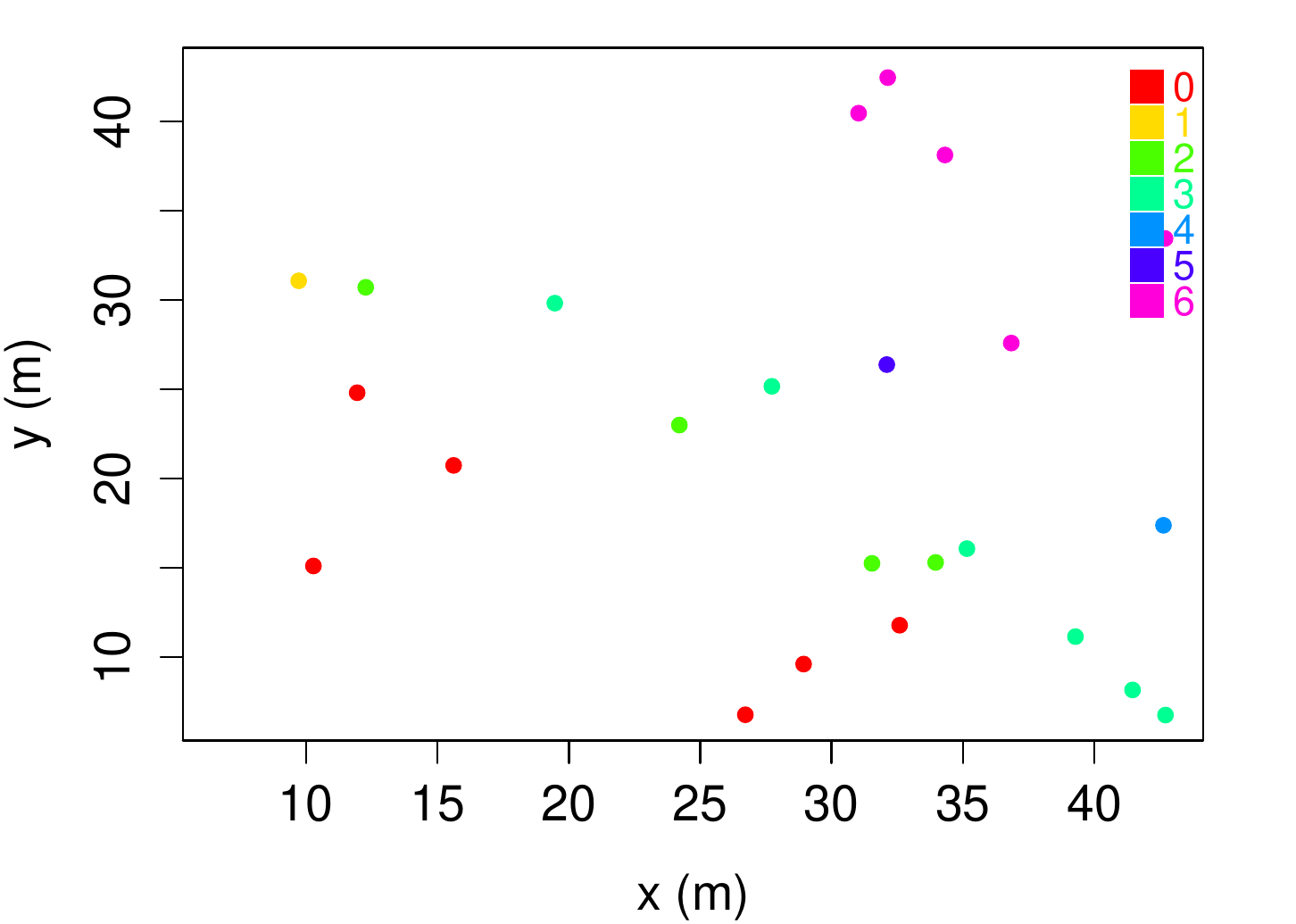}}
\caption{\it \small Example of scenario for multi-target tracking in binary sensors. We display the brute information extract from the sensors, which is an integer between $0$ and $N_t$.}
\label{fig:ex_mt}
\end{figure}

We provided in \cite{Ickowicz2008a} an accurate deterministic algorithm for the tracking of a target with a network of binary directional sensors. This algorithm used the $0-1$ information given by the sensors of the network to estimate the velocity direction, and then, the target position. Now, we consider that we have a \emph{known} number of targets $N_t$. However, the sensors are not able to make the association between the binary information and the targets, which means that our problem is equivalent to have a $0-N_t$ information for each sensor. The figure \ref{fig:ex_mt} provides an example of scenario and observation for this situation.

We assume that each target is independent from the others. The value of each sensor can be written,
\begin{eqnarray}
\label{eq:mt}
{\bf c}_i^{t} & = & \sum_{j=1}^N {\bf 1}_{\{ \langle {\bf x}_{t}^j-{\bf l}_{i}, {\bf v}_{t}^j \rangle <0 \}}
\label{eq:sg}
\end{eqnarray}
where ${\bf l}_i$ relates to the location of sensor $i$. This can also be viewed as the sum over the rows of the matrix $S_{t}$, 
\begin{equation}
\begin{array}{llll}
S_{t}  & = &  
\left(
\begin{array}{lllll}
0 & 1 & \hdots & 1 & 1 \\
0 & 0 & \hdots & 1 & 0 \\
\vdots & \vdots & \ddots & \vdots & \vdots \\
1 & 0 & \hdots & 1 & 1 \\
0 & 1 & \hdots & 0 & 0
\end{array}
\right)
\end{array}
,
\end{equation}
which is the binary matrix which row $i$ corresponds to the sensor $i$, and column $j$ stands for the target $j$. Therefore, estimating $S_t$ for all $t$ will provide the information needed to apply the algorithm of tracking presented in \cite{Ickowicz2008a}. From Figure \ref{fig:ex_mt}, we understand that with a huge number of sensor, we would be able to identify the separating lines, which would help for both target location and velocity estimation. Then, an appropriate association algorithm would help linking the targets to the identified lines.\\
Unfortunately, we assume that we just have a reasonable number of sensors (up to $50$, maximum). We presented in \cite{Ickowicz2013} an estimation procedure for $S_t$, using the importance sampling framework. Then, using the algorithm presented in \cite{Ickowicz2008a}, we could track several targets using binary sensors. The algorithm was efficient, but given the size of the parameter space ($N_s \times N_t$), the number of random sample to generate before reaching an accurate estimate is gigantic.\\
The present article provides a solution to overcome this dimension issue. The approach we use consists on using the Bayesian framework to provide a posterior distribution of the parameter of interest, $({\bfx}_t, {\bfv}_t)$. The modelling ideas are the following.
Given the observations $({\bfc}_t)$, we can derive the posterior distribution according to the following equation,
\begin{eqnarray}
p({\bfx}_t, {\bfv}_t \vert {\bfc}_t,{\bfx}_{t-1}, {\bfv}_{t-1}) & \propto & \underbrace{p({\bfc}_t \vert {\bfx}_{t-1:t}, {\bfv}_{t-1:t})}_{\textrm{likelihood}} \times \underbrace{p({\bfx}_t,{\bfv}_t \vert {\bfx}_{t-1}, {\bfv}_{t-1})}_{\textrm{prior}}
\end{eqnarray}
while the prior distribution raise directly from eq. \ref{eq:traj}, the likelihood need more work. All the information needed to draw ${\bfc}_t$ is brought by $({\bfx}_t,{\bfv}_t)$, therefore,
\begin{eqnarray}
p({\bfc}_t \vert {\bfx}_{t-1:t}, {\bfv}_{t-1:t}) &=& p({\bfc}_t \vert {\bfx}_{t}, {\bfv}_{t})
\end{eqnarray}
Now, if the number of target is equal to $1$, we can easily compute the likelihood of ${\bfc}_t$. Indeed, if we assume that $p_e$ is the probability that a sensor is giving the right $0-1$ information, and assuming the sensors are independent, we have,
\begin{eqnarray}
\label{eq:kl}
p({\bfc}_t \vert {\bfx}_{t}, {\bfv}_{t}) &=& \prod_i^{N_s} p(c_i^t \vert {\bfx}_t,{\bfv}_t) \nonumber \\
& = &  \prod_i^{N_s} \Big( \big[p_e {\bf 1}_{\mathcal{A}(i)} + (1-p_e) {\bf 1}_{\bar{\mathcal{A}}(i)}\big]^{c_i^t} \big[(1-p_e) {\bf 1}_{\mathcal{A}(i)} + p_e {\bf 1}_{\bar{\mathcal{A}}(i)}\big]^{1-c_i^t} \Big)\nonumber \\
\end{eqnarray}
where $\mathcal{A}(i) = \{ \langle {\bf x}_{t}-{\bf t}_{i}, {\bf v}_{t} \rangle <0 \}$, and $\bar{\mathcal{A}}(i)$ is its complementary.\\
If the number of target to track is $N_t > 1$, each $c_i^t$ is described in eq. \ref{eq:mt}.
And given that, we have no closed form expression for the distribution of a such variable as it only can be expressed as a sum of variables, giving,
\begin{eqnarray}
\label{eq:il}
p({\bfc}_t \vert {\bfx}_{t}, {\bfv}_{t}) &=&  \bigotimes\limits_{j=1}^{N_t} p({\bfc}_t \vert {\bfx}_t^j, {\bfv}_t^j)
\end{eqnarray}
which is intractable. Then, to perform an accurate multiple target tracking algorithm using the Bayesian approach, we have to work without the likelihood, or find a reasonable way to approximate it.
 Which we will perform in the next section.
 
\section{\ABc~Target Tracking}

\subsection{The \ABc~principle}
The first generalized approximate Bayesian computation algorithm aimed at provide some accurate enough inference with an approximation of the target density \cite{Pritchard1999}. This approach is widely used and popular among different areas such as medicine \cite{Sunnaker2013}, Several improvements have been proposed since, by improving the approximated likelihood with sequential methods \cite{Sisson2009, DelMoral2011, Sciences1999}, or empirical likelihood \cite{Mengersen2012}. The issue of possible poor acceptance rate is also adressed in \cite{Wegmann2009}, which offers a partial least square transformation to choose informative statistics.\\
The extension of the \ABc~algorithms provided by the SMC samplers, can be efficient but computer demanding. This limitation, thus, makes these extensions inefficient in our context given our real-time tracking requirements. Interested readers may also refer to \cite{Marin2011a}, which provides an excellent up-to-date survey of \ABc~methods.\\
In its most common form (the most very basic \ABc~algorithm is the approximate rejection algorithm (\ABcRej) which is detailed in the following section) \ABc~algorithms draw inference from the following posterior density,
\begin{eqnarray}
\pi_{\varepsilon} (\theta, \mathsf{x} \vert \mathsf{y}) &=& \frac{\pi(\theta) f(\mathsf{x} \vert \theta) {\bf 1}_{A_{\varepsilon,\mathsf{y}}}(\mathsf{x})}{\int_{A_{\varepsilon,\mathsf{y}}} \pi(\theta) f(\mathsf{x} \vert \theta) d\mathsf{x} d\theta}
\end{eqnarray}
where $\varepsilon$ is a tolerance level, $\rho$ an adequate distance function, and $A_{\varepsilon,\mathsf{y}}$ the set of "tolerated" observations, meaning,
\begin{eqnarray}
A_{\varepsilon,\mathsf{y}} &=& \{ \mathsf{x} : \rho(\mathcal{M}(\mathsf{x}),\mathsf{y}) < \varepsilon \}
\end{eqnarray}

\subsection{\ABc~Rejection Algorithm (\ABcRej)}

This algorithm is designed to overcome the issue of having intractable likelihood. It is the simplest one, but it is proved to perform exact inference in the presence of uniform model or measurement error \cite{Wilkinson2008}. The details are provided in Alg. \ref{alg:abc.rej}.\\

\begin{algorithm}
{The ABc rejection algorithm. \label{alg:abc.rej}}
\begin{algorithmic}[1]
\Procedure{\ABcRej}{$S,\rho$} \Comment{Need $S$ and $\rho$}
\For {t in $1:T$}
\State Draw $(\tilde{\bf x}_t, \tilde{ \bf v}_t)$ $\sim$ $\pi_0(\hat{\bf x}_{t-1},\hat{\bf v}_{t-1})$ \Comment{Sample using Eq. \ref{eq:traj}}
\State Compute $\tilde{\bfc}$ using $(\tilde{\bf x}_t, \tilde{\bf v}_t)$ \Comment{Apply $\mathcal{M}$ using Eq. \ref{eq:sg}}
\State Accept $(\tilde{\bf x}_t, \tilde{\bf v}_t)$ if $\rho(\tilde{\bf c},{\bf c}_t) < \varepsilon$ \Comment{Acceptance step (Eq. \ref{eq:d})}
\State Compute $(\hat{\bf x}_t,\hat{\bf v}_t) = 1/\vert a \vert (\sum \tilde{\bf x}_t^a,\sum \tilde{\bf v}_t^a)$ \Comment{Compute the estimates}
\EndFor
\EndProcedure
\end{algorithmic}
\end{algorithm}

The most important feature to have a successful \ABcRej~algorithm is to define a measure $\rho()$, between the model outputs, and to have a reasonable prior distribution. In our situation, we remind that the statistic $\bfc$ is a vector of length $N_s$, which values belong to $[0,N_t]$. We then simply define $\rho$ as,
\begin{eqnarray}
\rho({\bfc}_1,{\bfc}_2) &=& \| {\bfc}_1 - {\bfc}_2 \|_2^2
\label{eq:d}
\end{eqnarray}
the euclidean distance between the vectors ${\bfc}_1, {\bfc}_2$. 

\subsection{\ABc~Random Walk Algorithm (\ABcRW)}

This \ABcRej~algorithm provides a reasonable estimate of both position and velocity, but however doesn't really consider the past knowledge we have of the targets evolutions. Therefore, in addition to the Euclidian distance (Eq. \ref{eq:d}), we will also try to build a pseudo likelihood ($f$), to include the information we think relevant to this target tracking.

\begin{algorithm}
{The ABC Random Walk algorithm.\label{alg:abc.rw}}
\begin{algorithmic} [1]
\Procedure{\ABcRW}{$S,\rho$} \Comment{Need $S$ and $\rho$}
\For {t in $1:T$}
\For {i in $1:N_{part}$}
\State Draw $(\tilde{\bf x}_t^i, \tilde{\bf v}_t^i)$ $\sim$ $\pi_0(\hat{\bf x}_{t-1},\hat{\bf v}_{t-1})$ \Comment{Sample using Eq. \ref{eq:traj}}
\State Compute $\tilde{\bfc}^i$  using $(\tilde{\bf x}_t^i, \tilde{\bf v}_t^i)$ \Comment{Apply $\mathcal{M}$ using Eq. \ref{eq:sg}}
\State Generate $u \sim \mathcal{U}_{[0,1]}$
\State Accept $(\tilde{\bf x}_t^i, \tilde{\bf v}_t^i)$ \bf{if} 
\begin{eqnarray}
\rho(\tilde{\bfc}^i,{\bfc}_t) < \varepsilon & \textbf{and} & u \leq \frac{f(\tilde{\bf x}_t^i, \tilde{\bf v}_t^i) q(\tilde{\bf x}_t^{i-1}, \tilde{\bf v}_t^{i-1} \vert \tilde{\bf x}_t^{i}, \tilde{\bf v}_t^{i})}{f(\tilde{\bf x}_t^{i-1}, \tilde{\bf v}_t^{i-1}) q(\tilde{\bf x}_t^{i}, \tilde{\bf v}_t^{i} \vert \tilde{\bf x}_t^{i-1}, \tilde{\bf v}_t^{i-1})} \nonumber
\end{eqnarray}
\EndFor
\State Compute $(\hat{\bf x}_t,\hat{\bf v}_t) = 1/\vert a \vert (\sum \tilde{\bf x}_t^a,\sum \tilde{\bf v}_t^a)$
\EndFor
\EndProcedure
\end{algorithmic}
\end{algorithm}

Then, given that function $f$, we derive a classical MCMC step using the generation of a random uniform variable to select (or not) the new particles.\\
The function $f$ consist on a summary of all the information (a priori or based on the observations) that we think can help us improve the estimation procedure. Therefore, recalling that the main assumptions were based on a low-manoeuvring target, we will base $f$ on the consecutive bearings and distance between two consecutive measurements.

\subsection{Tuning Issues}

Usually, the acceptance rates of the \ABc-MCMC schemes are relatively low. In order to ensure a reasonable coverage of the target distribution, we therefore need to make sure that the acceptance rate stays at an acceptable level, $23\%$ being the heuristic most common when the parameter space dimension is bigger than $5$ \cite{Gelman1996}, \cite{Roberts1997}.\\
Improving the acceptance rate is usually a matter of prior distribution and renewal calibration (e.g. variance of the renewal process). As the chosen prior distribution seems rather adequate, we need to focus on the renewal process and on the \ABc~only features, namely $\rho$ and $\varepsilon$.\\
\paragraph*{Function of Statistics $\rho$} Given the nature of our output statistic (a vector with integer values), the choice of a Euclidean distance function appears reasonable. The results presented in section \ref{sec:results} let us think that this is the case.
\paragraph*{Tolerance level $\varepsilon$} This parameter will directly affect the acceptance rate, and increasing it will artificially increase this rate. The reason why we call it artificially is that a higher level will also lead to accept many unacceptable particle values. Therefore, to obtain the right targeted distribution, more particles will be needed. This situation is particularly inadequate for the purpose of live tracking. 
No, if we assume that no detection error can be made by the sensors, there is no reason to allow $\varepsilon$ to be above $1$ (because of the strict inequality). But this assumption being unrealistic, let assume a probability of error ($0$ instead of $1$) equal to $p_e$. Given the Binomial type distribution of the observations, a rough estimate of the error value per sensor would be $p_e \times N_t$. So, given a number of sensor $N_s$, we can choose $\varepsilon = N_s (p_e \times N_t)^2$.
\paragraph*{Renewal process} Bottom line, working on the renewal process appears as the best way to improve the acceptance rate, and hence, the posterior distribution. The renewal process we use in this article is the sampling from the prior distribution $\pi_0()$. Despite its apparent simplicity, it ensures that the main hypothesis for the Markov chain to converge are fulfilled.

\section{Advanced Features}

\subsection{Parallel Tempering}

The classical \ABc-MCMC schemes suffer form two majors drawbacks. First, the samples obtained are dependant, and often highly correlated. While this can be an issue for other features, there is nothing to worry about in our situation. Indeed, the transformation of the parameters (eq. \ref{eq:sg}) is not bijective, then the use of the \ABc-MCMC does not aim at drawing the true posterior distribution of $({\bfx}_t, {\bfv}_t)$. Instead, we expect that the acceptance step will select the most likely parameters according to the hypothesis we made on the movement (and represented by $f$ in \ABcRW). The second drawback may only be a concern if the targets are suspected to be highly manoeuvrable. In that situation, the \ABc-MCMC scheme offers no warranty to detect the changes of direction. To overcome this, we propose to use an \ABc~Parallel Temporing algorithm, as presented in \cite{Baragatti2012}.\\
The principle of parallel tempering is to introduce $N$ temperatures and to run in parallel $N$ associated MCMC chains with target distributions being tempered distributions of the target $\pi$. The first chain targets $\pi$, and the tempered chains target flatter versions of $\pi$. Each iteration of the parallel tempering is decomposed into two types of moves. Local moves, to update the chains. And global moves, to swap between the chains.\\
Transposed to \ABc~framework, the parallel tempering replaces the increasing temperatures with increasing tolerance values. The local moves are performed using the \ABc~framework. The global moves (or chains swap), demand a specific treatment.
Let $N$ be the size of the chain (also known as the number of particles), and $J$ be the number of parallel chains. Alg. \ABcPT~explain the additional step required to compute a parallel tempering algorithm.
\begin{algorithm}
{Parallel Tempering step of the ABC Parallel Tempering algorithm. \label{alg:abc.pt}}
\begin{algorithmic}[1] 
\Procedure{\ABcPT}{$(\pi^0_j)_j$}
\For {i in $1:N$}
\State Compute $({\bf x}_j, {\bf v}_j)_j$ according to \ABc-X.
\State Select $J$ pairs of chains in $[1,J]$ \Comment{subject to $j_l < j_l'$.}
\State Denote $i_1, j_1, \hdots, i_J, j_J$ the chosen pairs, with $i < j$.
\State Accept $({\bf x}_{i_l}, {\bf v}_{i_l}) \gets ({\bf x}_{j_l}, {\bf v}_{j_l})$ \bf{if} $\rho(\tilde{\bfc}_{j_l},{\bfc}_t) < \varepsilon_{i_l}$
\EndFor
\EndProcedure
\end{algorithmic}
\end{algorithm}
The issue with that kind of algorithm is the duration of the estimation process. Indeed, having $J$ chains means that your complexity increases drastically, and that is why this solution is accurate if you have highly turbulent targets.

\subsection{Reversible Jump}

We assumed so far that the number of targets is known. However, for the aim of surveillance in enemy areas, the exact number of targets to track is unknown, even if some knowledge of the range can be available. This issue has been studied quite widely for regular multi-target tracking, mostly using Particle Filters, by adding some prior information \cite{Boers2004}, proposing a two-layer algorithm \cite{Morelande2007,Yi2012}, or using a grid-based probabilistic approach. However, none of these methods can be applied in our particular context.\\
Having an evolving (or unknown) number of targets is an issue that rise a change of the parameter space dimension. This issue has been addressed in the context of MCMC by the use of reversible jump, introduced in \cite{Green1995}, and later improved in \cite{Green2009}. The idea is to calculate the likelihood for two different parameter number, and compute and acceptance step just like in the classical rejection algorithm.\\
This change in the parameter space can also be viewed as a model choice issue, and a review of \ABc~based model choice is provided in \cite{Toni2009a}. The techniques presented in this article however are very computer demanding in the sense that a lot of particles need to be drawn before a reasonable estimate is calculated. 

\section{Results}

\subsection{Single Target}

To underline the accuracy of our algorithms, we tested them on different simulation scenarios. First, we consider a single target and we compare the performances of the \ABc~algorithms with the MCMC Random Walk algorithm performed using the known likelihood (\ref{eq:kl}).
\begin{figure*}[ht!]
\centering
\includegraphics[height = 7cm, width = 12cm]{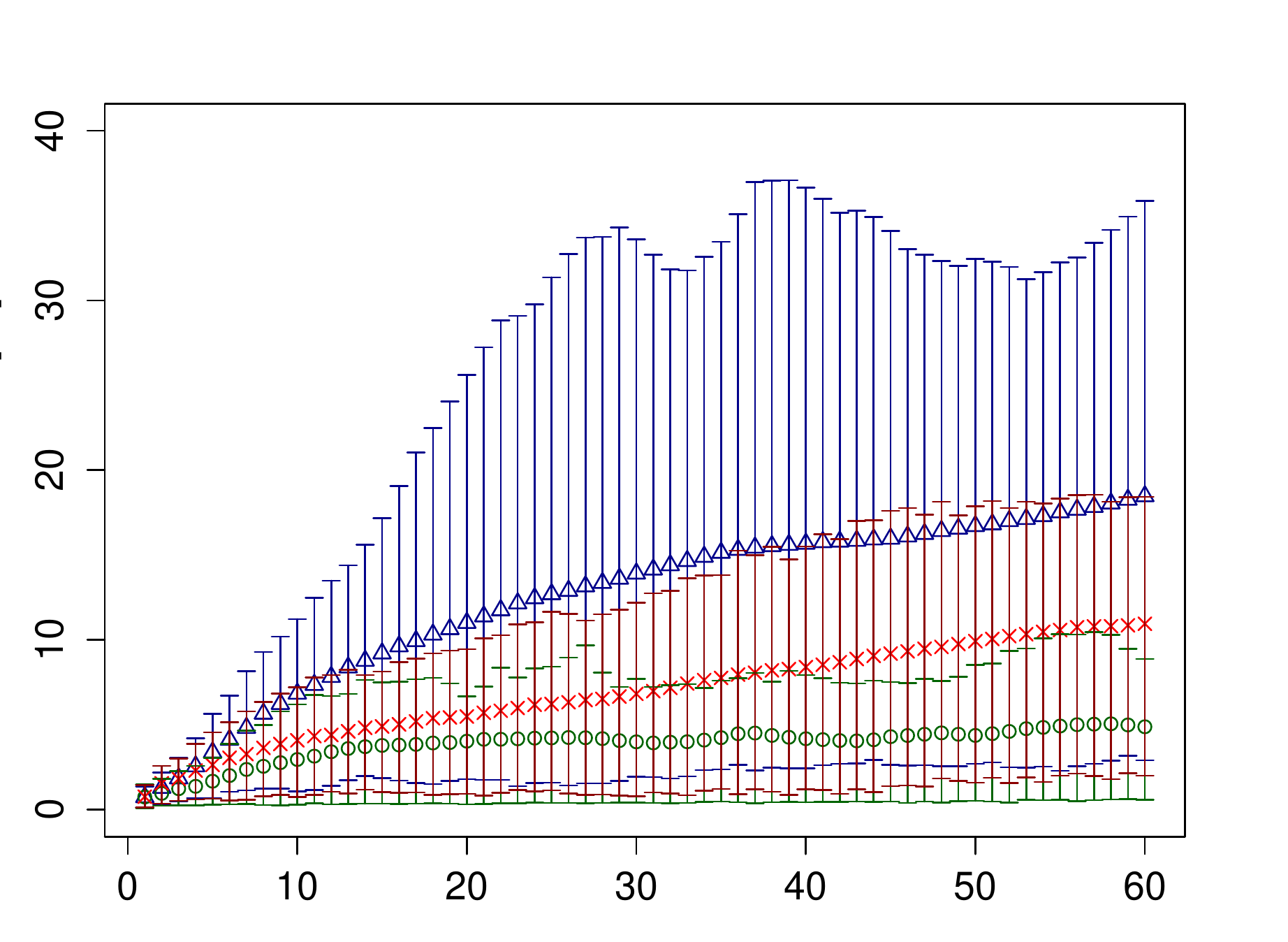}
\caption{\label{fig:rmse91} RMSE of the position estimation of $1$ target, for $64$ sensors. The green circles represents the MCMC algorithm, the red crosses the \ABcPT~algorithm,  and the blue triangles the \ABcRW~algorithm. We also plotted the $95\%$ confidence intervals for the MSE.}
\end{figure*}
We display on figure \ref{fig:rmse91} the evolution of the RMSE for the tracking of a single target, using the MCMC algorithm, the \ABcRW~algorithm, and the \ABcPT~algorithm. As expected, the MCMC performs better than the two others, the \ABcPT~algorithm being able to outperform the \ABcRW~algorithm. 

\subsection{Multiple Targets}

\begin{figure*}[ht!]
\centering
\includegraphics[height = 5.5cm, width = 6.5cm]{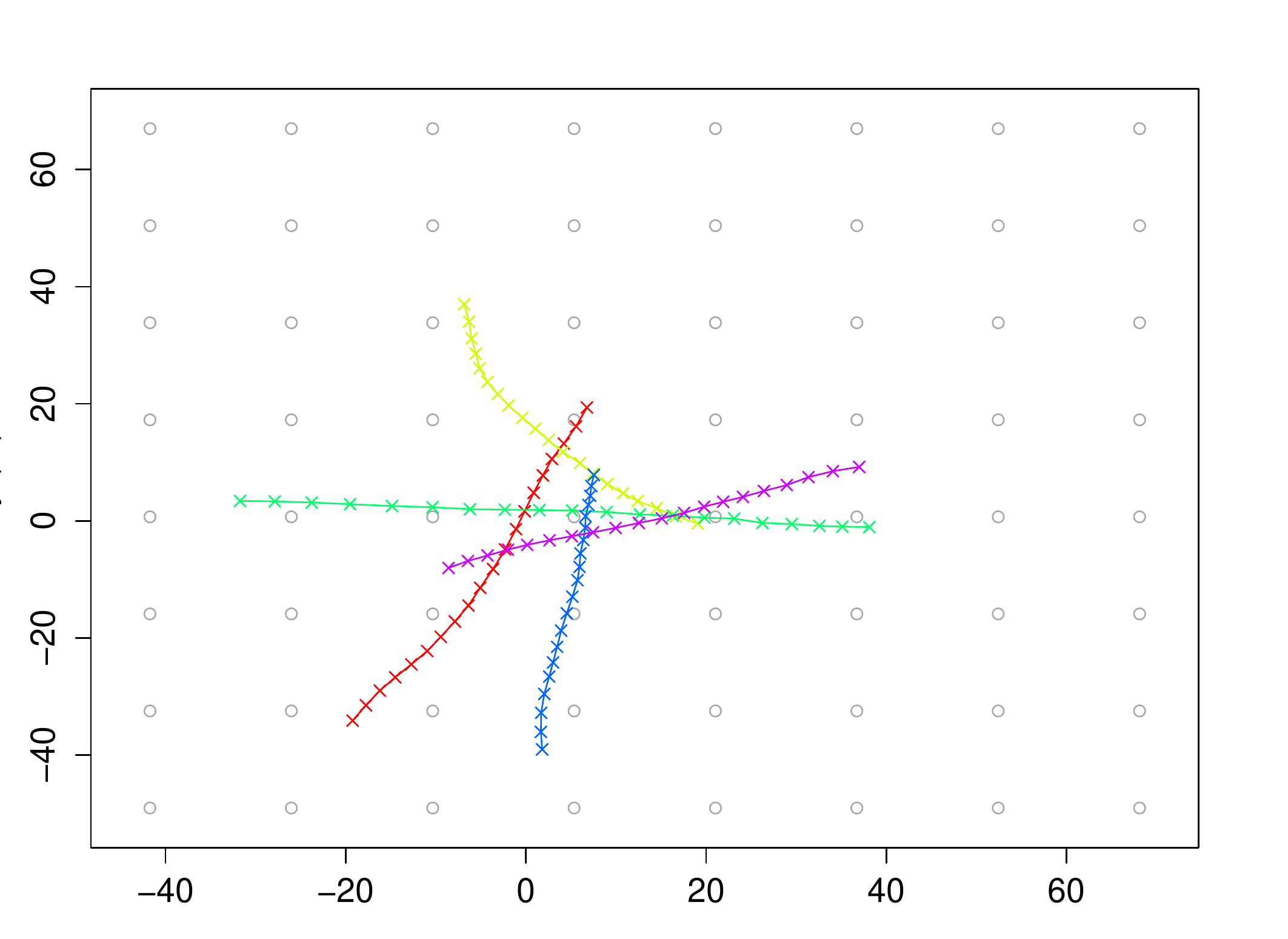}
\includegraphics[height = 5.5cm, width = 6.5cm]{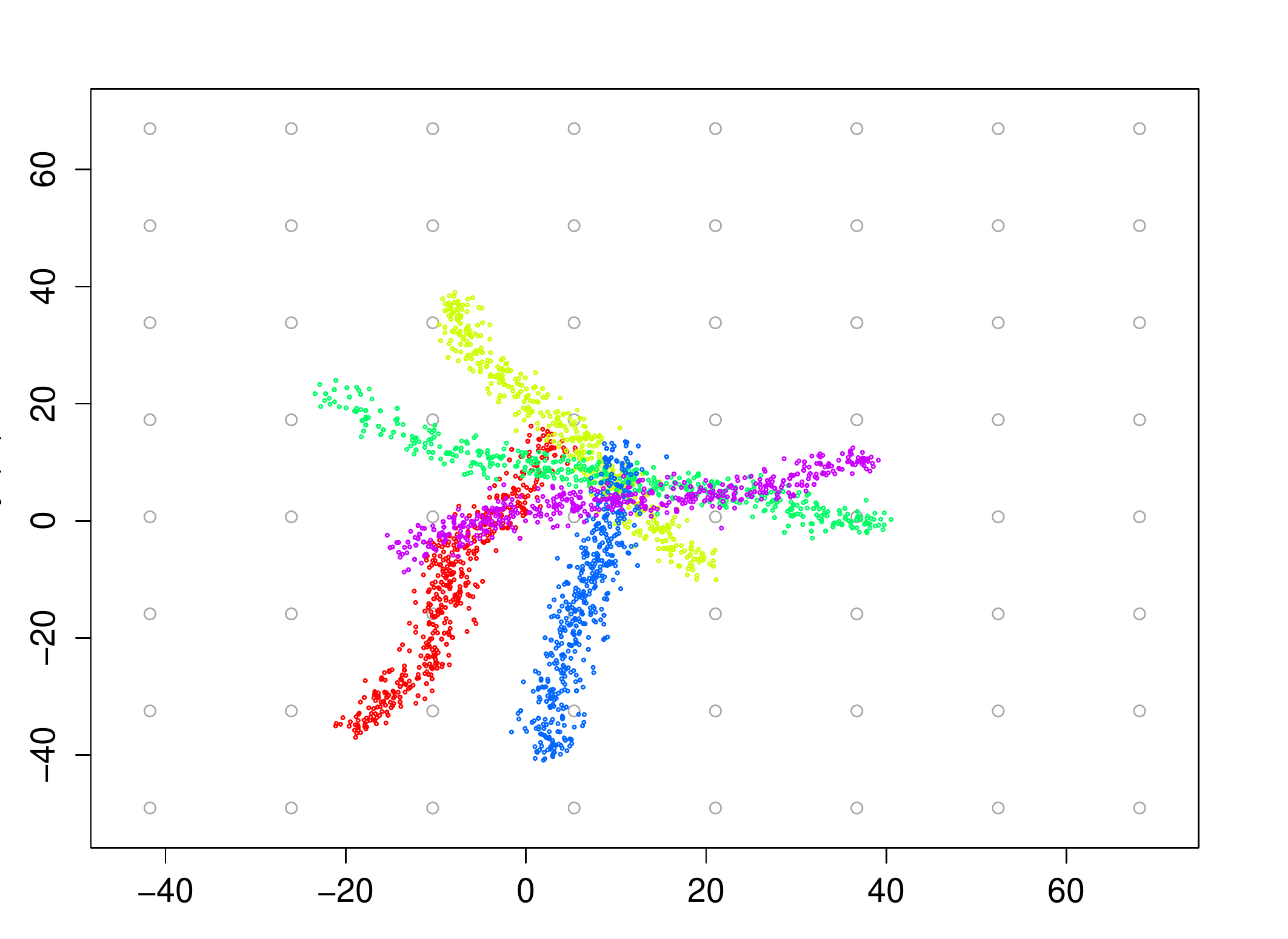}
\caption{\small \label{fig:5tt} Five target tracking scenario. The left panel represents the true trajectories, while the dots represent the \ABc~particles of the estimated trajectories.}
\end{figure*}

\begin{figure*}[ht!]
\centering
\includegraphics[height = 8cm, width = 12cm]{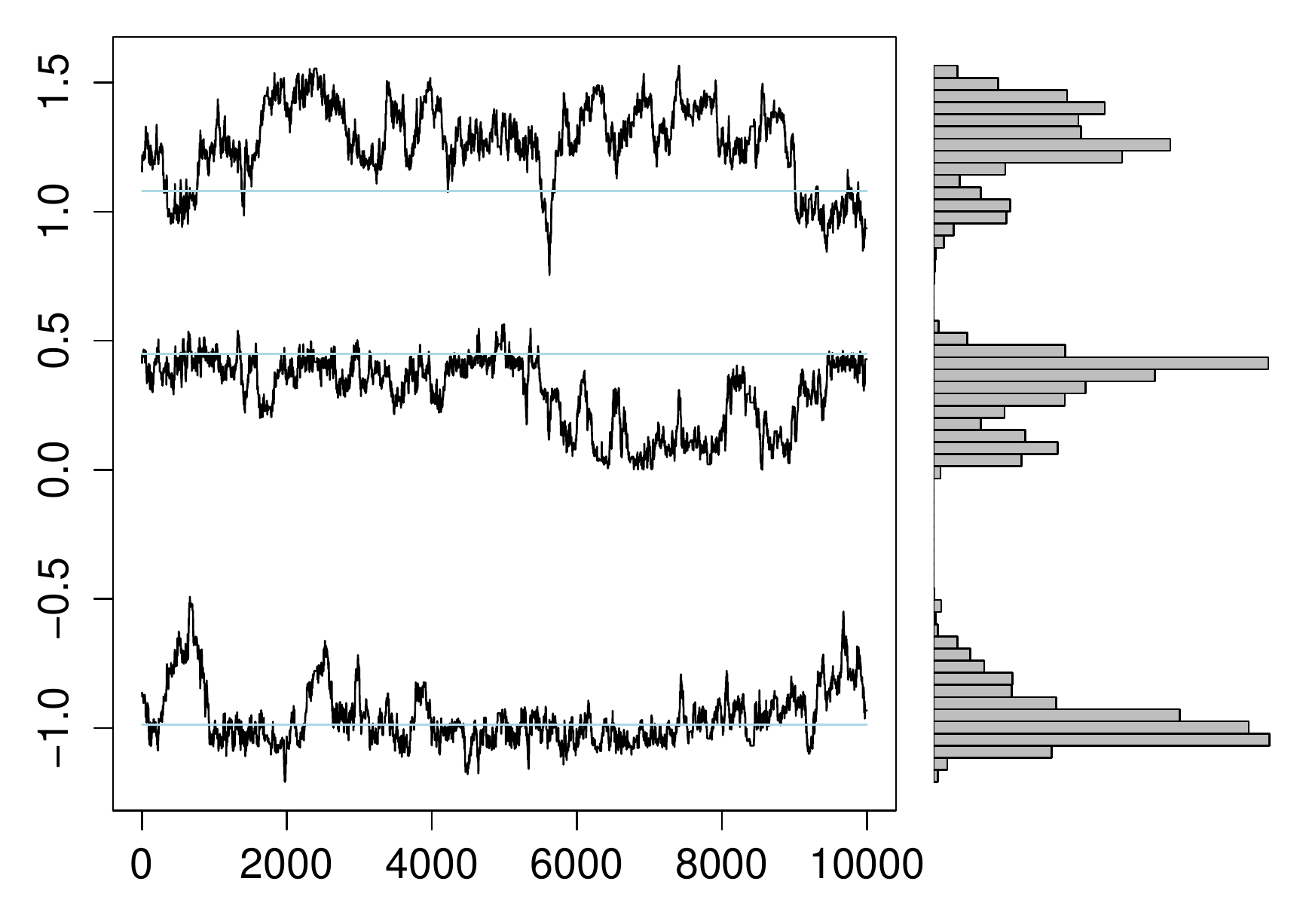}
\caption{\label{fig:rtv} Trajectories of the \ABcRW~algorithm, estimated bearings at one particular time $t$.}
\end{figure*}

\begin{figure*}[ht!]
\centering
\includegraphics[height = 9cm, width = 14cm]{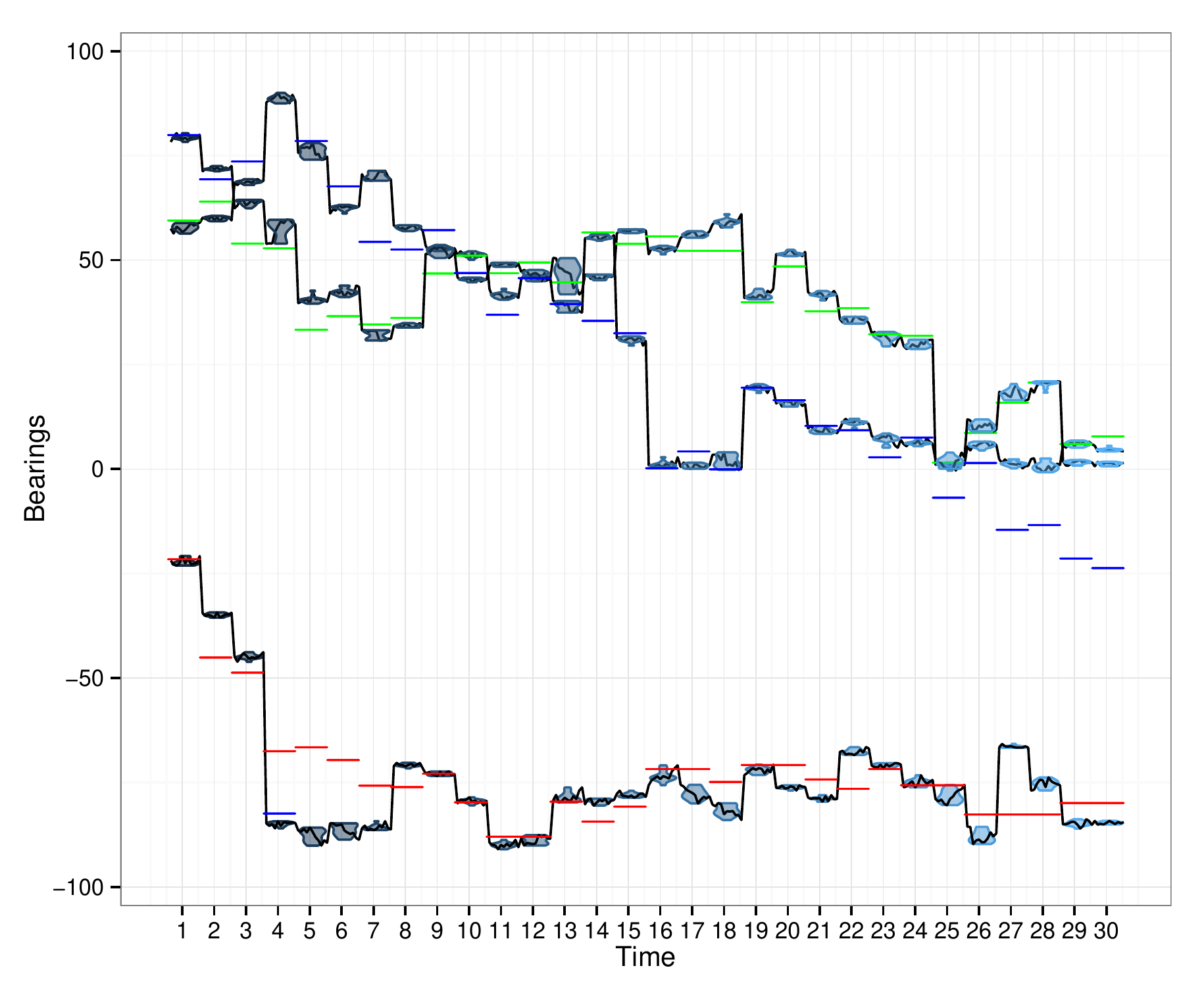}
\caption{\label{fig:rtp} Trajectories of the \ABcRW~algorithm,full path estimated bearings. The real bearings values are represented with different colors (green, blue, red).}
\end{figure*}

Then, we consider the trajectories of $N_t$ targets, which evolves according to equation \ref{eq:traj}, with $F_t = I_{N_t}$ and ${\bf Q}_t = \sigma^2 I_{N_t}$ (with $\sigma^2 = 0.1$). The starting positions of the target are randomly chosen on a circle centred in $0$, and with a radius equal to $40$. The initial velocities of the targets are chosen so that the targets move toward the center of the circle. The duration of the tracking is $30$ seconds. The simulations are computed through the statistical software $\mathsf{R}$.\\
On figure \ref{fig:5tt} we aimed at track $5$ targets, remarkably close to each other. The initial position of the targets was known within a range of $10$ meters. The MCMC runs where $5000$-particles long, and we display on this figure the last $250$, for each time period. The trajectories can be finally point-estimated using the Maximum A Posteriori (MAP). The non-turbulent nature of the trajectories ($\sigma_v = 0.1$) helped \ABc-RW~algorithm to provide a good estimation of all the targets trajectories.\\
\label{sec:results}
We display on figures \ref{fig:rtv} and \ref{fig:rtp} the particles accepted during the MCMC run, for a specific $t$ (fig. \ref{fig:rtv}), or the whole trajectory (fig. \ref{fig:rtp}). The simulation consist on $3$ targets, with more varying directions than in the previous simulation ($\sigma_v = 0.3$). The total number of particles in a run is $10000$. The acceptance rate is roughly $0.12$, which is a bit low, but due to the choice of $\varepsilon$. 
In figure \ref{fig:rtv}, we also present the histograms estimating the posterior density of the bearings. As we can notice from both figures, our algorithm is able to track the three targets with a high accuracy, which is remarkable since no association rule is defined.\\
Finally, we evaluated the root mean square error (RMSE) of the estimated position for different number of targets and different number of sensors. Table \ref{tble:MSE_MTT} provides the different RMSE values and their variance as the tracking is being performed for $30$ seconds. The obvious conclusion is that the best algorithm seems to be the one using the Parallel Tempering feature (\ABcPT). However, we recall that the estimation has been performed with $5$ parallel chains, which means $5$ times more computer resources. While it has no importance for offline tracking, real-time tracking is a different issue and the \ABcRW~algorithm provides good estimates as well.
\begin{landscape}
\begin{table*}[ht!]
\centering
\begin{tabular}{cc|ccc|ccc}
  & & \multicolumn{3}{c}{16} & \multicolumn{3}{c}{64}\\
  & & \ABcRej & \ABcRW & \ABcPT &  \ABcRej & \ABcRW & \ABcPT\\
     \hline
     \hline
    $N_t : 2$ & $t : 10$ & 9.70  (3.40)&  8.80  (3.32)&  5.34 (1.18)& 9.96 (3.57)&  9.27 (2.56)&  3.39 (1.44)\\
              & $t : 20$ & 20.37 (6.20)& 18.96  (6.17)& 8.29 (4.81)& 21.64 (6.71)& 17.24 (6.12)&  6.03 (5.59)\\
              & $t : 30$ & 34.21 (9.83)& 21.98 (6.89)& 12.75 (6.06)& 37.32 (10.37)& 21.73 (6.89)& 9.39 (5.64)\\
              \hline
    $N_t : 3$ & $t : 10$ & 11.88 (6.02)&  9.30 (3.78)&  6.84 (1.56)&  12.91 (4.26)&  9.87 (3.39) & 3.15 (1.42)\\
              & $t : 20$ & 20.53 (7.07)& 19.96 (6.88)& 12.50 (4.72)&  20.15 (5.21)& 19.11 (6.56) & 5.34 (4.28)\\
              & $t : 30$ & 36.91 (11.05)& 23.30 (6.27)& 15.88 (6.40)&  36.93 (11.63)& 23.06 (6.86) & 13.25 (5.45)\\
              \hline
    $N_t : 4$ & $t : 10$ & 11.62 (3.18)&  9.64 (3.20)&  6.02 (1.48)& 11.25 (3.82)& 10.23 (3.26)&  4.80 (2.23)\\
              & $t : 20$ & 25.35 (8.14)& 20.90 (7.79)& 12.67 (3.00)& 24.58 (8.41)& 20.39 (8.00)&  12.57 (4.43)\\
              & $t : 30$ & 43.85 (14.84)& 23.31 (7.07)& 16.80 (5.56)& 43.18 (14.22)& 24.12 (6.96)& 16.24 (6.40)\\
  \hline
 \hline
\end{tabular}
\caption{\label{tble:MSE_MTT} Final RMSE of the location estimator of $N_t$ targets, with different number of sensors ($16$ and $64$). The RMSE is calculated at $t=10$s, $t=20$s and $t=30$s, with $300$ particles. The standard error of the RMSE is specified under brackets.}
\end{table*}
\end{landscape}
\section{Conclusion}

The recent developments in low-power wireless sensors, with cheaper prices, made them necessary in several fields, and especially for commercial and military applications. In this paper, we considered a sensor network made with two main components, the sensors and the centralized fusion center. The objects of interest are targets evolving in a watched area, and the fusion center should combine the binary informations provided by the sensors to obtain the trajectories estimations.

The constraints on the targets' trajectories are minimum, and the observations consist on binary directional measurements. We have built an estimation of the targets' trajectories based on the Approximate Bayesian computation principle by considering our model with the inverse problem framework. Then, using different features, we provide a complete algorithm that can be used for an unknown number of manoeuvrable targets.\\
This whole new algorithm does however still lack of softness to address the tuning issues. This is particularly true for the choice of $\rho$, which is of prime importance for the quality of the estimation. Adopting the adaptive approach from the literature \cite{Sciences1999}, \cite{DelMoral2011} could be helpful, but could raise some computing-time issues, especially for the purpose of real time tracking. An analyse of the influence of the network topology could also be helpful in identifying the robustness flows.

\bibliographystyle{IEEEtran}
\bibliography{/home/ickowicz/Documents/Donnees/Articles/Bibtex/Publications-ABC-based_MTT.bib}

\end{document}